%% file: Lat2018proc.tex
\documentclass{PoS_twospeakers}

\title{Towards the P-wave nucleon-pion scattering amplitude in the $\Delta (1232)$ channel}

\ShortTitle{Towards the P-wave nucleon-pion scattering amplitude}
\author{{\speaker{Srijit Paul}{Srijit Paul and Giorgio Silvi}}$\;^{ab}$, {\speaker{Giorgio Silvi}{Srijit Paul and Giorgio Silvi}}$\;^{d}$, Constantia Alexandrou$\;^{ac}$, Giannis Koutsou$\;^{a}$, Stefan Krieg$\;^{bd}$, Luka Leskovec$\;^{e}$, Stefan Meinel$\;^{fg}$, John W. Negele$\;^{h}$, Marcus Petschlies$\;^{i}$, Andrew Pochinsky$\;^{h}$, Gumaro Rendon$\;^{f}$ and Sergey Syritsyn$^{j}$\\
\llap{$^a$}Computation-based Science and Technology Research Center, The Cyprus Institute \\
20 Kavafi Str., Nicosia, 2121, Cyprus\\
\llap{$^b$}Faculty of Mathematics und Natural Sciences, University of Wuppertal\\
Wuppertal-42119, Germany\\
\llap{$^c$}Department of Physics, University of Cyprus, POB 20537, 1678 Nicosia, Cyprus\\
\llap{$^d$}Institute for Advanced Simulation, Forschungszentrum J\"ulich GmbH, J\"ulich 52425, Germany\\
\llap{$^e$}Theory Center, Jefferson Lab, Newport News, VA 23606, USA\\
\llap{$^f$}Department of Physics, University of Arizona, Tucson, AZ 85721, USA\\
\llap{$^g$}RIKEN BNL Research Center, Brookhaven National Laboratory, Upton, NY 11973, USA\\
\llap{$^h$}Center for Theoretical Physics, Massachusetts Institute of Technology\\
Cambridge, MA 02139, USA\\
\llap{$^i$}Helmholtz-Institut f{\"u}r Strahlen- und Kernphysik, University of Bonn, Bonn 53115, Germany\\
\llap{$^j$}Department of Physics and Astronomy, Stony Brook University, Stony Brook, NY 11794, USA\\
E-mail: \email{s.paul@hpc-leap.eu}, \email{g.silvi@fz-juelich.de}, \email{alexand@ucy.ac.cy}, \email{koutsou@cyi.ac.cy}, \email{s.krieg@fz-juelich.de}, \email{leskovec@jlab.org}, \email{smeinel@email.arizona.edu}, \email{negele@mit.edu}, \email{marcus.petschlies@hiskp.uni-bonn.de}, \email{avp@mit.edu}, \email{jgrs@email.arizona.edu},  \email{ssyritsyn@quark.phy.bnl.gov} }

\include{abstract}

\FullConference{The 36th Annual International Symposium on Lattice Field Theory - LATTICE2018\\
		22-28 July, 2018\\
		Michigan State University, East Lansing, Michigan, USA.}
\usepackage{gensymb}
\usepackage{bbm}
\usepackage{amsmath}
\usepackage{longtable}
\usepackage{hhline}
\usepackage{booktabs}
\usepackage{fixmath}
\usepackage{verbatim}
\usepackage{color, colortbl}
\definecolor{Gray}{gray}{0.9}
\graphicspath{{images/}{./}}

\newcommand{\Nucleon}{N}

\begin{document}


\input{Introduction}


\input{LatticeSetup}


\input{Interpolators}


\input{MovingFrames}


\input{AngularMomentum}


\input{ProjectionMethod}


\input{Spectra}


\input{ScatteringAnalysis}


\input{Phase_shift_results}


\input{Discussion}


\input{Acknowledgements}






\providecommand{\href}[2]{#2}\begingroup\raggedright\endgroup

\end{document}

%% file: abstract.tex
\abstract{
  We use lattice QCD and the L\"uscher method to study elastic pion-nucleon scattering in the isospin $I = 3/2$ channel,
  which couples to the $\Delta(1232)$ resonance. 
  Our $N_f=2+1$ flavor lattice setup features a pion mass of $m_\pi \approx 250$ MeV,
  such that the strong decay channel $\Delta \rightarrow \pi N$ is close to the threshold.
  We present our method for constructing the required lattice correlation functions from single- and two-hadron
  interpolating fields and their projection to irreducible representations of the relevant symmetry group of the lattice. We show preliminary results for the energy spectra in selected moving frames and irreducible
  representations, and extract the scattering phase shifts. Using a Breit-Wigner fit,
  we also determine the resonance mass $m_\Delta$ and the $g_{\Delta-\pi N}$ coupling.
}

%% file: Introduction.tex
\section{Introduction}

The study of scattering of strongly-interacting hadrons on the lattice has given us quantitative theoretical insight into unstable hadrons, which are otherwise difficult to analyze perturbatively \cite{Stoler:1991vi}. Even though the theoretical foundations for studying the $2\rightarrow 2$ scattering resonances on the lattice were laid in 1990s by L\"uscher \cite{L_scher_1990}, the generalizations \cite{Rummukainen_1995,Kim_2005,leskovec2012scattering, Briceno:2014oea}, further extensions \cite{G_ckeler_2012,Briceno:2017tce},  practical implementation of these algorithms along with the development of solvers(multigrid) and the availability of lattice ensembles generated using the physical pion mass took two decades. In the last decade, there have been extensive studies of low-lying meson resonances, starting with the $\rho$ meson \cite{Alexandrou:2017mpi,Gottlieb:1985rc,McNeile:2002fh,Aoki:2007rd,Gockeler:2008kc,
Jansen:2009hr, Feng:2010es, Frison:2010ws,Lang:2011mn,Aoki:2011yj,Pelissier:2012pi,
Dudek:2012xn,Wilson:2015dqa, Bali:2015gji,Bulava:2016mks,Hu:2016shf,Guo:2016zos,Fu:2016itp,Andersen:2018mau}
that served as the first evidence for the practical applicability of the L\"uscher methodology for extracting resonance parameters. The next step is to study more complex $2\rightarrow 2$ strongly interacting scattering systems. Here we explore the nucleon-pion scattering in the $I=3/2$ and $J^P=3/2^+$ channel where the lowest-lying baryon resonance, the $\Delta(1232)$, is located. This resonance has a mass of $\approx 1210$ MeV and a decay width of $\Gamma_{\Delta\to N\pi} \approx 117$ MeV \cite{Olive:2016xmw}. The $I=3/2$ $P$-wave $N\pi$ channel is the dominant decay mode, with a branching fraction of $99.4\%$. The PDG only lists one other decay mode - $N\gamma$ with a branching fraction $0.6\%$. The process is almost completely elastic \cite{shrestha2012multichannel}, but nearby inelastic resonances with similar quantum numbers could have a small contribution on the phase shift that needs to be taken into account in the analysis.

The lattice computation involves the evaluation of two-point correlation functions between the QCD-stable interacting fields: pion, nucleon, and the interpolating field with the quantum numbers of the resonance. The correlators are built from a combination of smeared forward, sequential and stochastic propagators. An important step in the proper identification of the spectra comes from the use of the projection method. It is used to build interpolators that transform according to a given irrep $\Lambda$ and row $r$, and overlap to the angular-momentum eigenstates of interest.

Previous studies of the $\Delta$ coupling to the $N\pi$ channel have used the Michael-McNeile method to determine the coupling \cite{Alexandrou:2013ata} as well as the L\"uscher method \cite{Meissner:2010ij,Verduci:2014btc,Andersen:2017una}. None of the previous studies have fully taken into account the partial-wave mixing. In this preliminary work, we also neglect this mixing, but we plan to include it in the full-fledged calculation.


%% file: LatticeSetup.tex
\section{Lattice setup}

Our calculation is based on the gauge-field ensembles 
with the actions and algorithms described in Ref.~\cite{Durr:2010aw}.
The gluon action is the tree-level improved Symanzik action, and the same clover-improved Wilson
action is used for the sea and valence quarks. The gauge links in the fermion action are smeared
using two levels of HEX smearing \cite{Durr:2010aw}.

Here we report on our preliminary results from $3072$ measurements on the \texttt{A7} ensemble, whose parameters
are given in Table \ref{tab:lattice}. This constitutes the first part of our two-part study of pion-nucleon scattering at $m_{\pi} \approx 250$ MeV.
The parameters of the ensemble \texttt{A8} for the follow-up simulation with larger spatial volume 
are listed in the table as well.


\begin{table}
\begin{center}
\begin{tabular}{|c|cccccc|}
\hline
Label          & $N_s^3\times N_t$ & $\beta$ & $a m_l$    & $a\:(\rm fm)$     & $m_\pi$ (MeV)  & $m_\pi L$\cr
\hline\hline
\texttt{A7}  & $24^3\times 48$   & $3.31$  & $-0.09530$  & $ 0.116$   & $        258.3(1.1)$   & $3.6$\cr
\texttt{A8}  & $32^3\times 48$   & $3.31$  & $-0.09530$  & $ 0.116$   & $\approx 250$   & $4.8$\cr
\hline
\end{tabular}
\end{center}
\caption{Parameters of the lattice gauge-field ensembles \texttt{A7} (presently used) and \texttt{A8}.}
\label{tab:lattice} 
\end{table}

%% file: Interpolators.tex
\section{Interpolators and correlation functions}
\label{seq:Interpolators}

To extract the lattice spectrum in the nucleon and $\Delta$ channel, we construct correlation matrices from
two-point functions of 1- and 2-hadron interpolating fields with quantum numbers of the nucleon ($\Nucleon$) and Delta ($\Delta$), respectively.
As building blocks we use the familiar single-hadron interpolating fields for the pion, nucleon and $\Delta$. In the following,
$\vec{p}$ denotes the three-momentum, and we omit the time coordinate for brevity.

The pion interpolator is given by
\begin{equation}
\pi(\vec{p})=\sum_{\vec{x}} \bar{d}(\vec{x}) \gamma_5 u(\vec{x}) e^{i\vec{p}\cdot\vec{x}}\,.
\end{equation}

\begin{figure}
\centering
\includegraphics[width=0.43\textwidth]{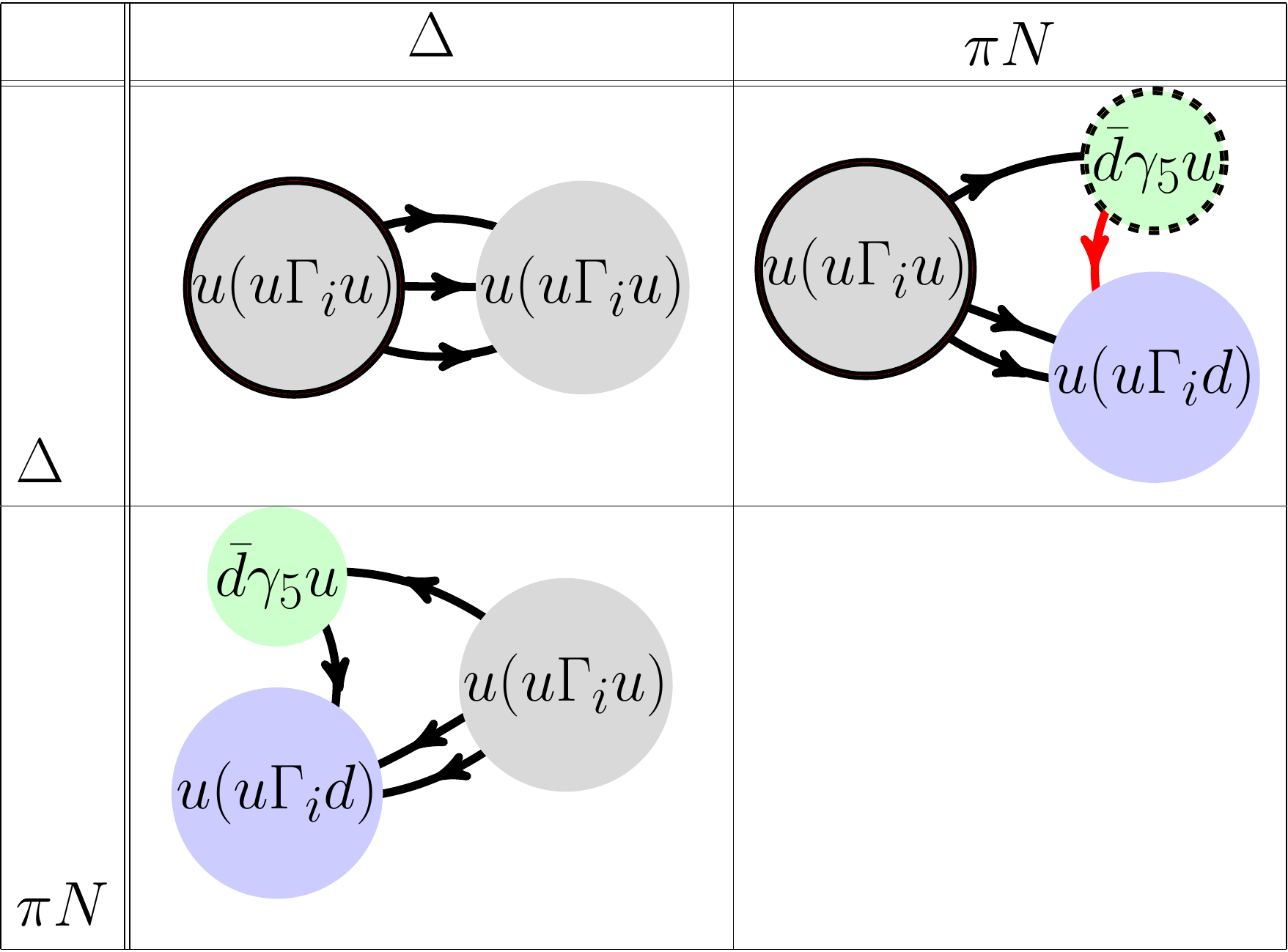}%
\hfill
\includegraphics[width=0.37\textwidth]{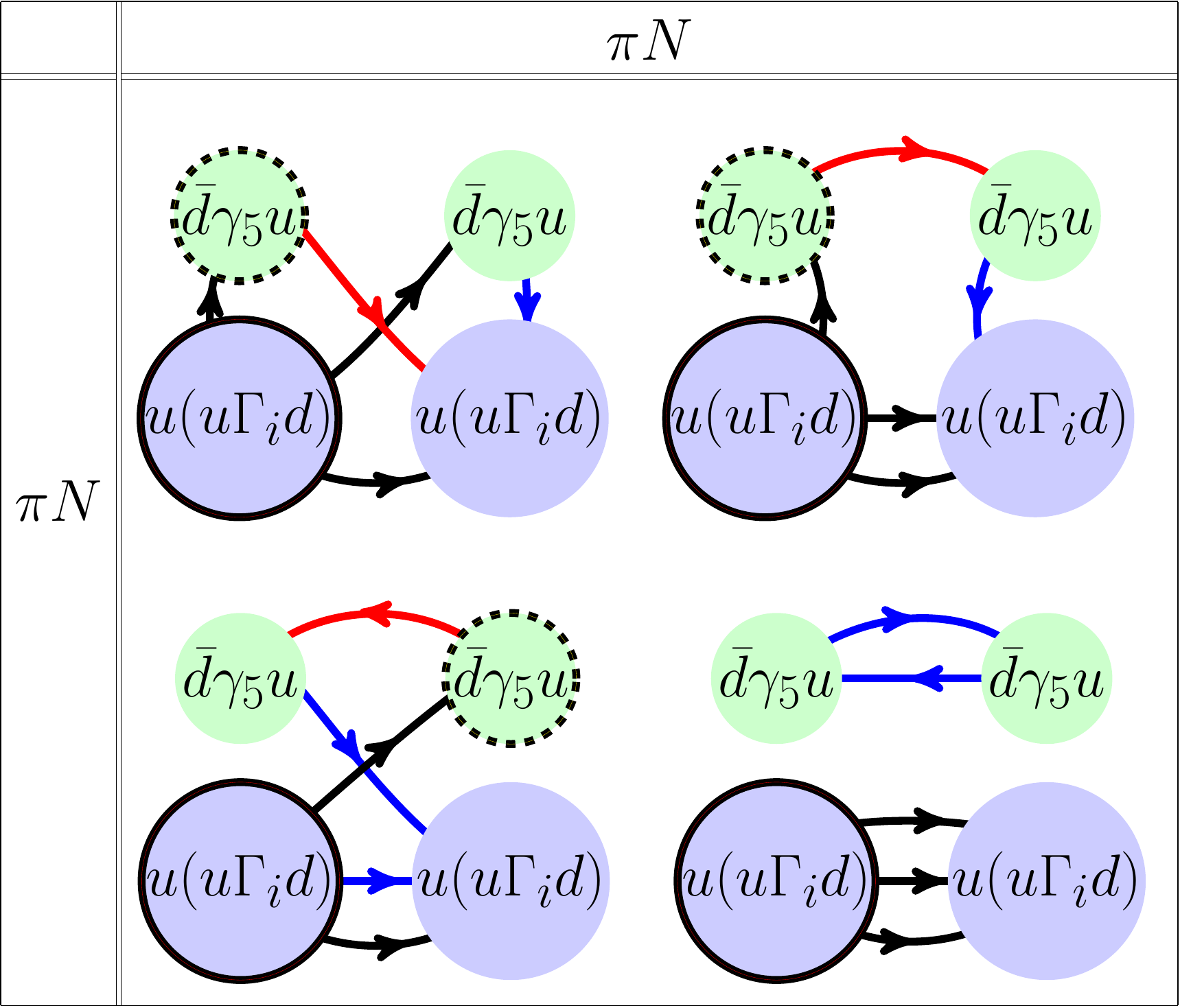}
\caption{Left panel: Two-point function contractions involving the $\Delta$ interpolator. Grey circles represent the $\Delta$, green circles represent the $\pi$, and blue circles represent the $N$. A circle with a red outline represents a point source, while the dotted outline represents a sequential source. The black arrow lines represent point-to-all propagators, and the red arrow lines represent sequential propagators. Right panel: Two-point function contractions for $\pi N - \pi N$, otherwise analogous to the left panel. The blue arrow lines represent stochastic propagators.}
\label{fig:contract1}
\end{figure}

For both the nucleon and the Delta, we include two choices of interpolators,
\begin{equation}
\begin{split}
\Nucleon_\mu^{(1)} (\vec{p})= \sum_{\vec{x}} \epsilon_{abc} 
( u_{a}(\vec{x}))_\mu ( u_b^T(\vec{x})C \gamma_5 d_c(\vec{x})) \,  e^{i\vec{p}\cdot\vec{x}}, \\
\Nucleon_\mu^{(2)} (\vec{p})= \sum_{\vec{x}} \epsilon_{abc} 
(u_{a}(\vec{x}))_\mu ( u_b^T(\vec{x})C\gamma_0\gamma_5 d_c(\vec{x})) \, e^{i\vec{p}\cdot\vec{x}}\,,
\end{split}
\end{equation}
and
\begin{equation}
\begin{split}
\Delta_{\mu i}^{(1)} (\vec{p})= \sum_{\vec{x}} \epsilon_{abc} 
( u_{a}(\vec{x}))_\mu ( u_b^T(\vec{x})C \gamma_i u_c(\vec{x}))\,  e^{i\vec{p}\cdot\vec{x}}, \\
\Delta_{\mu i}^{(2)} (\vec{p})= \sum_{\vec{x}} \epsilon_{abc} 
( u_{a}(\vec{x}))_\mu ( u_b^T(\vec{x})C \gamma_i \gamma_0 u_c(\vec{x}))\,  e^{i\vec{p}\cdot\vec{x}}.
\end{split}
\end{equation}
Finally, the two-hadron interpolators are obtained from products of the form
\begin{equation}
\Nucleon_\mu^{(1,2)} (\vec{p}_{N})\,\pi(\vec{p}_{\pi}).
\end{equation}
The two-point functions obtained from the interpolators above are evaluated by Wick contraction 
and factorization into products of diagram building blocks.
The latter are calculated numerically from point-to-all, sequential and stochastic propagators.
Figure \ref{fig:contract1} shows the topology types for the diagrams obtained from the $\Delta - \Delta$ and $\Delta - \pi N$ 
two-point correlation functions in the left panel. The right panel shows the same for the $\pi N - \pi N$ correlator.
In both panels, each topology represents 2 to 4 actual diagrams.

%% file: MovingFrames.tex
\section{Moving frames}

In order to apply the L{\"u}scher method we have to locate our energy region of interest below the $N\pi\pi$ threshold.
Due to the quantization of momenta, $\vec{p} = \frac{2\pi}{L}(n_x, n_y, n_z)$, and the fairly small volume of ensemble \texttt{A7}, 
the energy spectrum in the $\pi N$ or $\Delta$ rest frame is quite sparse below $E_{\mathrm{thr}} = m_{N} + 2 m_{\pi}$,
which leaves only a few energy values located in the region of interest.

In order to better constrain the phase shift curve with additional points, we make use of moving frames.
The Lorentz boost contracts the box, giving a different effective value of the spatial length $L$ along the boost direction \cite{leskovec2012scattering}. 
Our choices of total momenta for the $\pi N$ system are listed in the first column of Table \ref{plan}.
 
  \begin{table}[!htb]
  \centering
\resizebox{\textwidth}{!}{%
\begin{tabular}{| l | l | l | l | l | l |}
\hline
$\frac{L}{2\pi}\vec{P}_{ref} [N_{dir}] $ & Group & $N_{elem}$ & $\Lambda(J):\pi ($ $0^-$ $) $& $\Lambda(J):N ($ $\frac{1}{2}^+$ $)$& $\Lambda(J):\Delta ($ $\frac{3}{2}^+$ $)$ \\
\hline \hline &        &         &       &                   &                                                                  \\
$(0,0,0)$  $[1]$ & $O_{h}^D$  & 96 & $A_{1u}($ $0$ $,4,...)$ & $G_{1g}($ $ \frac{1}{2}$ $,\frac{7}{2},...)\oplus G_{1u}($ $ \frac{1}{2}$ $,\frac{7}{2},...)$ & $H_{g}($ $\frac{3}{2}$ $,\frac{5}{2},...) \oplus H_{u}($ $\frac{3}{2}$ $,\frac{5}{2},...)$ \\
&        &         &       &                   &                                                                  \\
$(0,0,1)$  $[6]$ & $C_{4v}^D$ & 16 & $A_{2}($ $0$ $,1,...)$ & $G_{1}($ $ \frac{1}{2}$ $,\frac{3}{2},...)$ & $G_1(\frac{1}{2},$ $\frac{3}{2}$ $,...) \oplus G_2($ $\frac{3}{2}$ $,\frac{5}{2},...) $\\ 
&        &         &       &                   &                                                                  \\
$(0,1,1)$  $[12]$ &$C_{2v}^D$ & 8 & $A_{2}($ $0$ $,1,...)$ &  $G($ $\frac{1}{2}$ $,\frac{3}{2},...)$ & $G(\frac{1}{2},$ $\frac{3}{2}$ $,...)$\\ 
&        &         &       &                   &                                                                  \\
$(1,1,1)$ $[8]$ & $C_{3v}^D$ & 12 & $A_{2}($ $0$ $,1,...)$ &  $G($ $ \frac{1}{2}$ $,\frac{3}{2},...)$ & $G(\frac{1}{2},$ $\frac{3}{2}$ $,...) \oplus F_1($ $\frac{3}{2}$ $,\frac{5}{2},...) \oplus F_2($ $\frac{3}{2}$ $,\frac{5}{2},...)$\\
&        &         &       &                   &                                                                  \\
\hline
\end{tabular}
}
\caption{Choices of total momenta (and numbers of equivalent directions), along with the relevant symmetry groups, irreducible representations $\Lambda$, and their angular momentum content.}
\label{plan}
\end{table}

%% file: AngularMomentum.tex
\section{Angular momentum on the lattice}

In the continuum, states are classified according to their angular momentum $J$, corresponding to the irreducible representations of $SU(2)$, and parity $P$.
On the lattice, the rotational symmetry is broken and the symmetry group is reduced to the double-cover of the cubic group \cite{johnson1982angular,basak2005group}. The 48 elements of the cubic group $O_h$ (rotation and inversion operations) leave the lattice grid unchanged. 
Since we characterize states of spin-1/2 and spin-3/2 fields in this work,
we need to consider the double-cover group $O_h^D$, which includes the $2\pi$ rotations (negative identity), thus having twice the number of elements \cite{Bernard:2008ax}.

In moving frames, the symmetry is further reduced to smaller groups (\textit{little groups}), which have a smaller set of irreps and a higher mixing of angular momenta in each one of them. Additionally, parity does not represent a useful quantum number in boosted frames, since both parities are featured in the same irrep. In Table \ref{plan} we list the symmetry groups associated with the moving frames, along with their irreps containing the angular momenta of interest. This many-to-one mapping of angular-momentum states leads to angular-momentum mixing on the lattice. 
In most irreps, we must determine not only the $( J=3/2, l=1) $ amplitude where the $\Delta$ appears as a resonance,
but also the $( J=3/2, l=2)$, $(J=1/2,l=0 )$ and $( J=1/2,l=1 )$ amplitudes 
that could have a small contribution from nearby resonances (listed in Table \ref{resonances}). Fortunately, these remain non-resonant in the energy region below approximately $1.6$ GeV.

\begin{table}
\centering
\begin{tabular}{|l | c | c|}
\hline
                Particle & $J^P$ & $\Gamma_{N\pi}[MeV]$\\ \rowcolor{Gray}
                \hline \hline
        $\Delta(1232)$ & $3/2^+ $ & $112.4(5)$\\        \hline
        $\Delta(1600)$ & $3/2^+ $ & $18(4)$\\
        $\Delta(1620)$ & $1/2^- $ & $37(2)$\\
        $\Delta(1700)$ & $3/2^- $ & $36(2)$\\
        \dots              & \dots      & \dots\\
        \hline
\end{tabular}
\caption{The $\Delta(1232)$ and nearby resonances with listed $J^P$ and decay width in $N\pi$ \cite{shrestha2012multichannel}.}
\label{resonances}
\end{table}

%% file: ProjectionMethod.tex
\section{Projection method}

We apply group-theoretical projections to obtain operators that transform irreducibly under rotations and reflections of the proper symmetry group. 
The master formula to project to a specific irrep $\Lambda$, row $r$ and occurrence $n$ is given by \cite{Morningstar:2013bda}
\begin{equation}
O^{\Lambda,r,n}(p)= \frac{d_\Lambda}{g_{G^D}}\sum_{R \in G^D} \Gamma_{r,r}^\Lambda (R) U_{R} \phi(p) U^{\dagger}_{R}  \;\;\;\;\; r \in \{1,\dots ,d_\Lambda\},
\label{pm}
\end{equation}
where $\Gamma^{\Lambda}$ are the representation matrices of the elements $R$ (rotations and reflections) of the double group $G^D$, $\phi(p)$ is a single/multi hadron operator (e.g.,~the operators presented in Sec.~\ref{seq:Interpolators}), $U_{R}$ is the quantum field operator that applies the transformation $R$. Additionally, $d_\Lambda$ is the dimension of the irrep $\Lambda$, and $g_{G^D}$ is the number of elements in the group $G^D$.

\begin{figure}
\centering
\includegraphics[width=.7\textwidth]{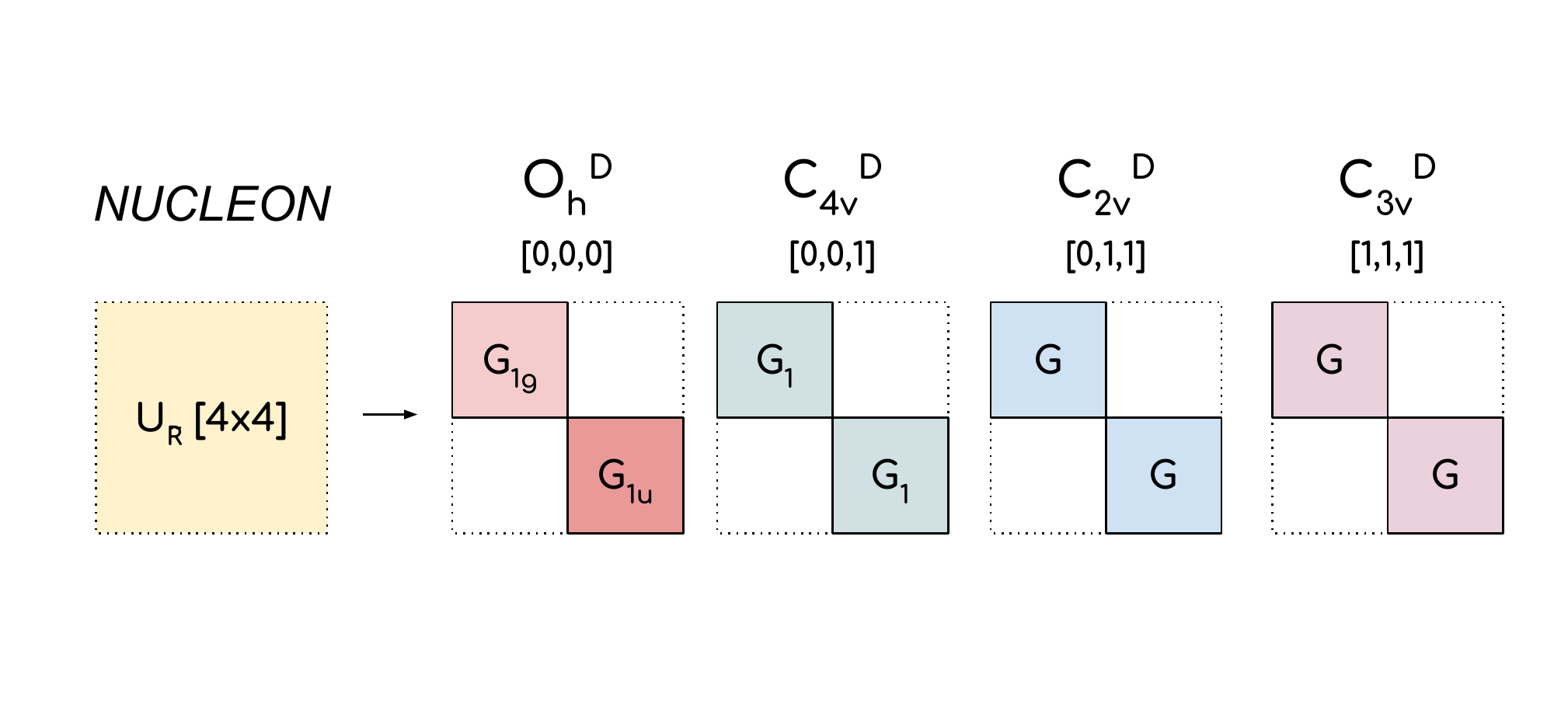}
\caption{ Tensor decomposition of the nucleon transformation matrices. Applying Eq.~(\ref{occ}) to the nucleon transformation matrices $S(R)$ tells us the occurrence of the irreps containing $J=1/2$ in each frame. This information guides us in the projection of the single-nucleon interpolator. 
It shows that in the rest frame we can have one projected interpolator for each of the irreps $G_{1g}$ and $G_{1u}$.
On the other hand, in each moving frame, we can build two linearly independent nucleon interpolators since the same irrep has a double occurrence.}
\label{Fig:nt}
\end{figure}

\begin{figure}
\centering
\includegraphics[width=.7\textwidth]{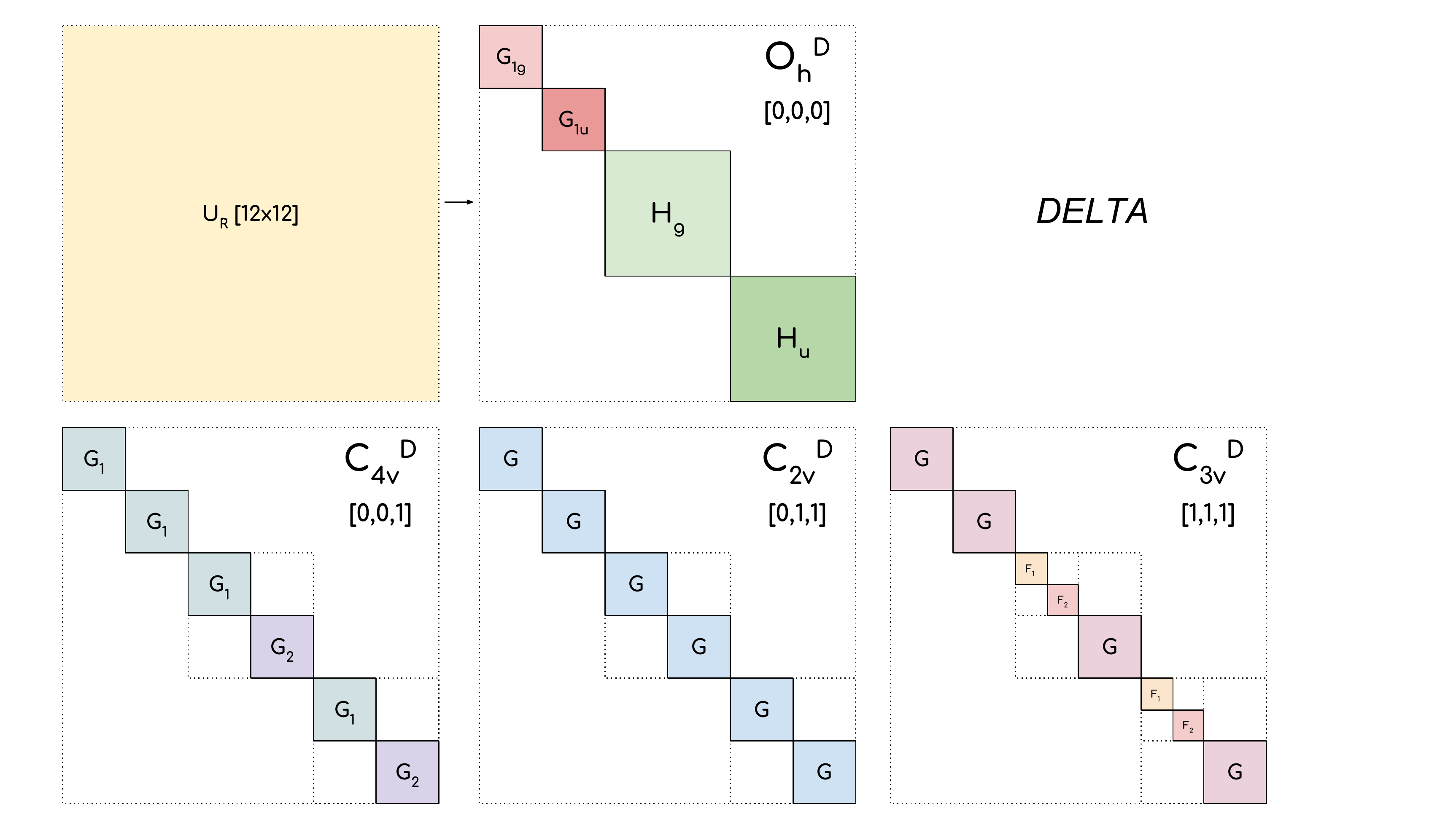}
\caption{ Tensor decomposition of the Delta transformation matrices. 
  The transformation matrix for the Delta comes from the tensor product $A(R)\otimes S(R)$ in \ref{rotops} (with inversion from \ref{invops}). 
  Applying Eq.\ref{occ} shows the occurrence of irreps containing $J=3/2$. 
  In the rest frame there are only single occurrences, while in moving frames there are multiple occurences. 
  For example, for the group $C_{2v}^D$ the same irrep $G$ appears $6$ times. In fact, we can build $6$ independent projected operators for a single $\Delta$ in this irrep.  }
\label{Fig:dt}
\end{figure}

\begin{table}
\begin{center}
\begin{tabular}{|c | c | c | c | c |c|}
\hline
         $Irr \backslash J$    & $J=1/2$ & $J=3/2$ & $J=5/2$& $J=7/2$& $J=9/2$\\
\hline\hline 
		 $G_{1 g/u}$ & 1 & 0 & 0 & 1 & 1   \\
	     $G_{2 g/u}$ & 0 & 0 & 1 & 1 & 0\\
	     $H_{g/u}$   & 0 & 1 & 1 & 1 & 2 \\ 
	     \hline
\end{tabular}
\caption{\label{spincont}Occurrences of irreps of the group $O_h^D$ (rest frame) in the subduction of half-integer-$J$ irreps of $SU(2)$. The subscripts $g/u$ indicate the parity \textit{gerade/ungerade} (even/odd).} 
\end{center}
\end{table}

The single-hadron operators transform under rotations as
\begin{equation}
\begin{split}
R\pi(t,\vec{x})R^{-1}&=\pi(t,R^{-1}\vec{x}), \\
R\Nucleon(t,\vec{x})R^{-1}&=S(R)\Nucleon(t,R^{-1}\vec{x}), \\
R\Delta(t,\vec{x})^\alpha_k R^{-1}&=A(R)_{kk^\prime}S(R)\Delta_{k^\prime}(t,R^{-1}\vec{x}), \\
\end{split}
\label{rotops}
\end{equation}
where $A(R)=U^1(\omega,\Theta,\Psi)$ denotes the 3-dimensional irrep of $SU(2)$, and $S(R)$ is the 2-dimensional spinor representation of $SU(2)$, 
\begin{gather} 
S(R)=
\begin{bmatrix}
    U^{1/2}(R)   & 0 		  \\
    0            & U^{1/2}(R) 
\end{bmatrix}.
\end{gather}
The inversions are given by
\begin{equation}
\begin{split}
I\pi(t,\vec{x})I^{-1}&=-\pi(t,-\vec{x}),\\
I\Nucleon(t,\vec{x})I^{-1}&=\gamma_t\Nucleon(t,-\vec{x}), \\
I\Delta(t,\vec{x})I^{-1}&=\gamma_t\Delta(t,-\vec{x}).
\end{split}
\label{invops}
\end{equation}
Our choice of the Euclidean $\gamma$-matrices is the DeGrand-Rossi basis.

The choice of total momentum $\vec{P}$ determines the relevant symmetry group. Looking at Table \ref{spincont}, the $J$ value of the hadron tells us in which irrep (of the rest frame) it should be contained. In order to have a complete identification of the irreps, we deduce them from the characters $\chi$ (trace) of the transformation matrices. It is possible to find the occurrence $m$ of the irrep $\Gamma^\Lambda$ in the matrix $M(R)$ realizing the transformation (e.g. $S(R)$ for the single nucleon) using the formula \cite{moore2006multiparticle,cotton2003chemical}
\begin{equation}
m=\frac{1}{g_{G^D}}\sum_{R \in G^D} \chi^{\Gamma^\Lambda (R)}\chi^{M(R)}.
\label{occ}
\end{equation}
The multiplicities $m$ give us the numbers of occurrences of the irreps for the specific operator we want to project (see Table \ref{spincont}).
This corresponds to the number of independent projected operators we can extract for a specific irrep $\Lambda$ and row $r$. In Figs.~\ref{Fig:nt} and \ref{Fig:dt}, we show schematically the decomposition of the transformation matrices $S(R)$ for the nucleon and $A(R)\otimes S(R)$ for the Delta in all frames relevant for this study.

Once we correctly identified the tensor decomposition in each irrep, we used our code to project the $N$, $\Delta$ and $N \pi$ interpolators (the single $\pi$ does not need projection). Most of the projections lead to linearly dependent operators \cite{Morningstar:2013bda}. 
We made use of the Gram-Schmidt procedure to construct linear combinations of operators orthogonal to each other.
As an example for the nucleon-pion system, we show one projected operator in the rest frame in irrep $H_g$ and row $r=1$:
\small
\begin{equation}
\frac{1}{2} \pi (1,0,0) \Nucleon_2(-1,0,0)+\frac{1}{2} i\pi(0,1,0)\Nucleon_2(0,-1,0)-\frac{1}{2} i\pi(0,-1,0)\Nucleon_2(0,1,0)-\frac{1}{2} \pi (-1,0,0). \Nucleon_2(1,0,0).
\nonumber
\end{equation} 
\normalsize The momentum directions are given in brackets and the subscript of the nucleon operator labels the Dirac index. 
Taking into account all linearly independent operators, rows and momentum directions in the 8 irreps of the 4 frames considered,
we reach a total 1720 projected operators for Delta and the nucleon-pion system. 
In order to maximize the statistics, we use all of them when building the correlation functions.

%% file: Spectra.tex

\section{Spectrum results}

\begin{figure}
\centering
\includegraphics[width=.48\textwidth]{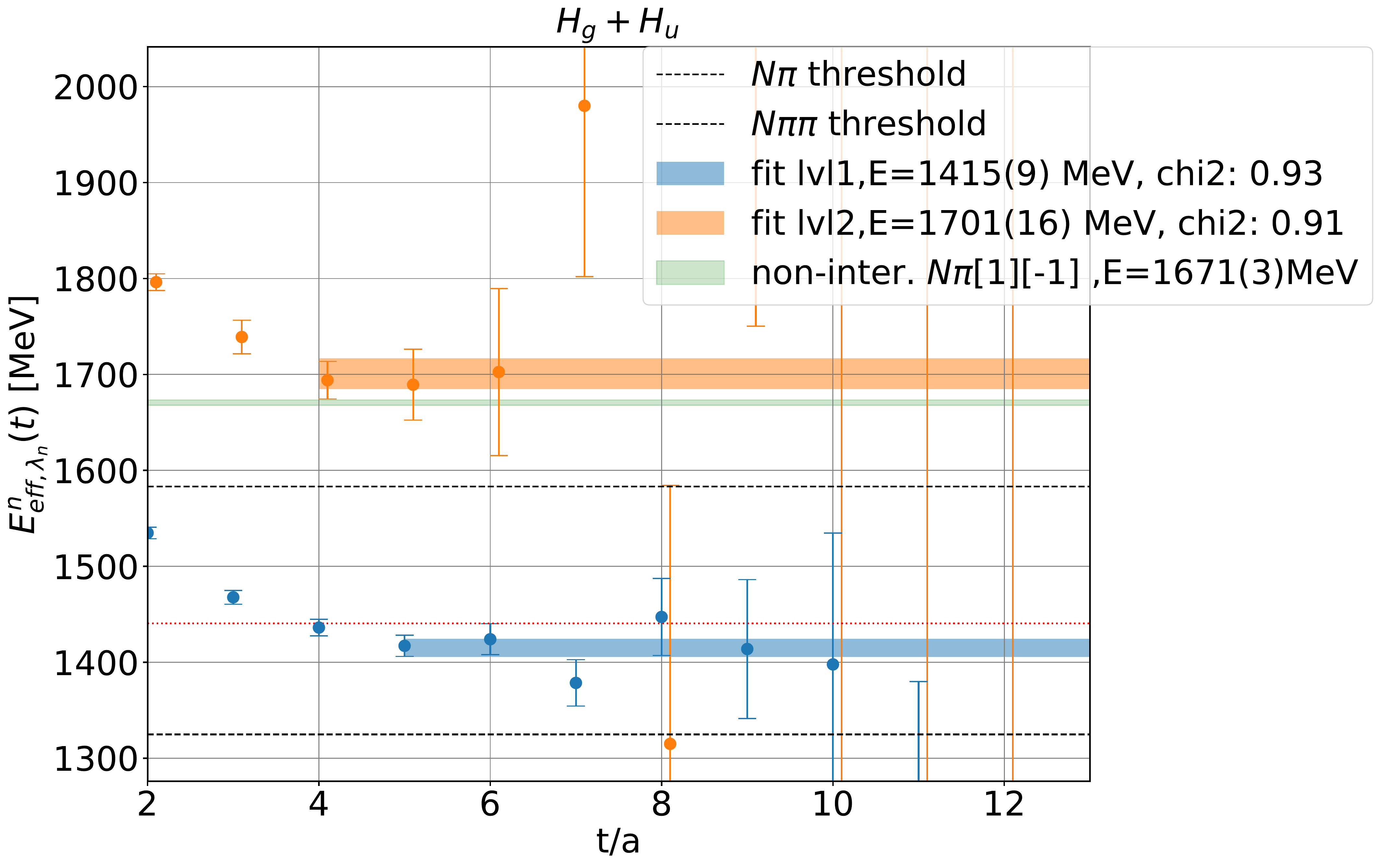}
\includegraphics[width=.48\textwidth]{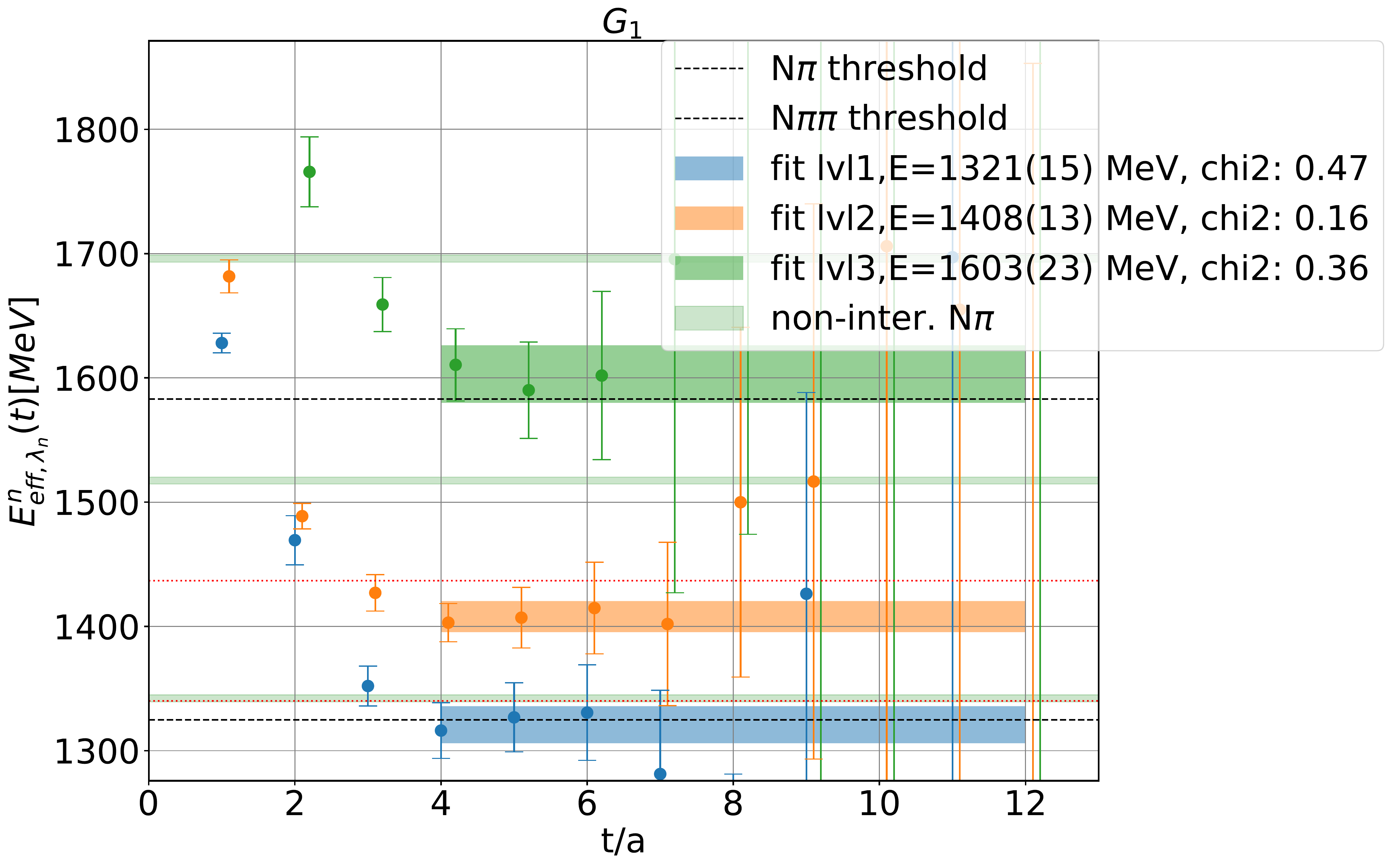}
\caption{\label{Fig:emp}Effective-energy plots. Left panel: Rest frame $\vec{P}=\frac{2\pi}{L}(0,0,0)$: group $O_h^D$, irreps $H_g$ (forward-propagating) and $H_u$ (backward-propagating). The ground state has maximum overlap with the single-hadron $\Delta$-like interpolator, while the first excited level shows an expected shift in energy with respect to the first non-interacting nucleon-pion energy. Right panel: Moving frame $\vec{P}=\frac{2\pi}{L}(0,0,1)$: group $C_{4v}^D$, irrep $G_1$. The ground state has dominant overlap to the $N\pi$ two-hadron interpolator, while the second energy level overlaps dominantly with the $\Delta$ operator. The other two levels display a shift in energy with respect to their non-interacting counterparts, in the direction away from the resonance.}
\end{figure}

With a basis of projected interpolators we construct correlation matrices $C_{ij}^{\Lambda,r,m}$ 
and make use of the variational method (Generalized EigenValue Problem) \cite{Blossier:2009kd} to determine the energy spectrum in each irrep.
Figure \ref{Fig:emp} shows the effective energies and the fit results for the chosen time ranges.
We perform single-exponential fits directly on the principal correlators.

\begin{figure}
\centering
\includegraphics[width=.48\textwidth]{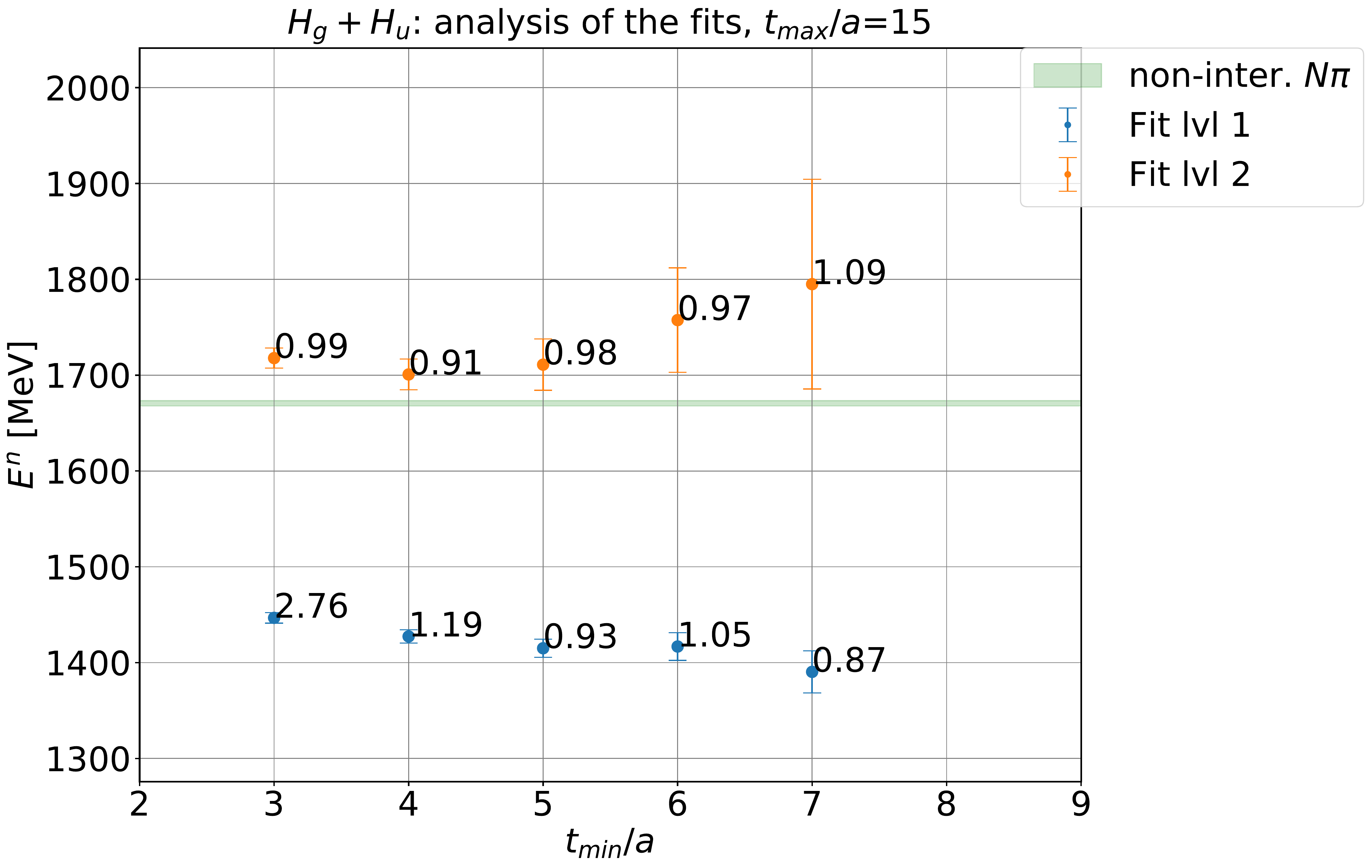}
\includegraphics[width=.48\textwidth]{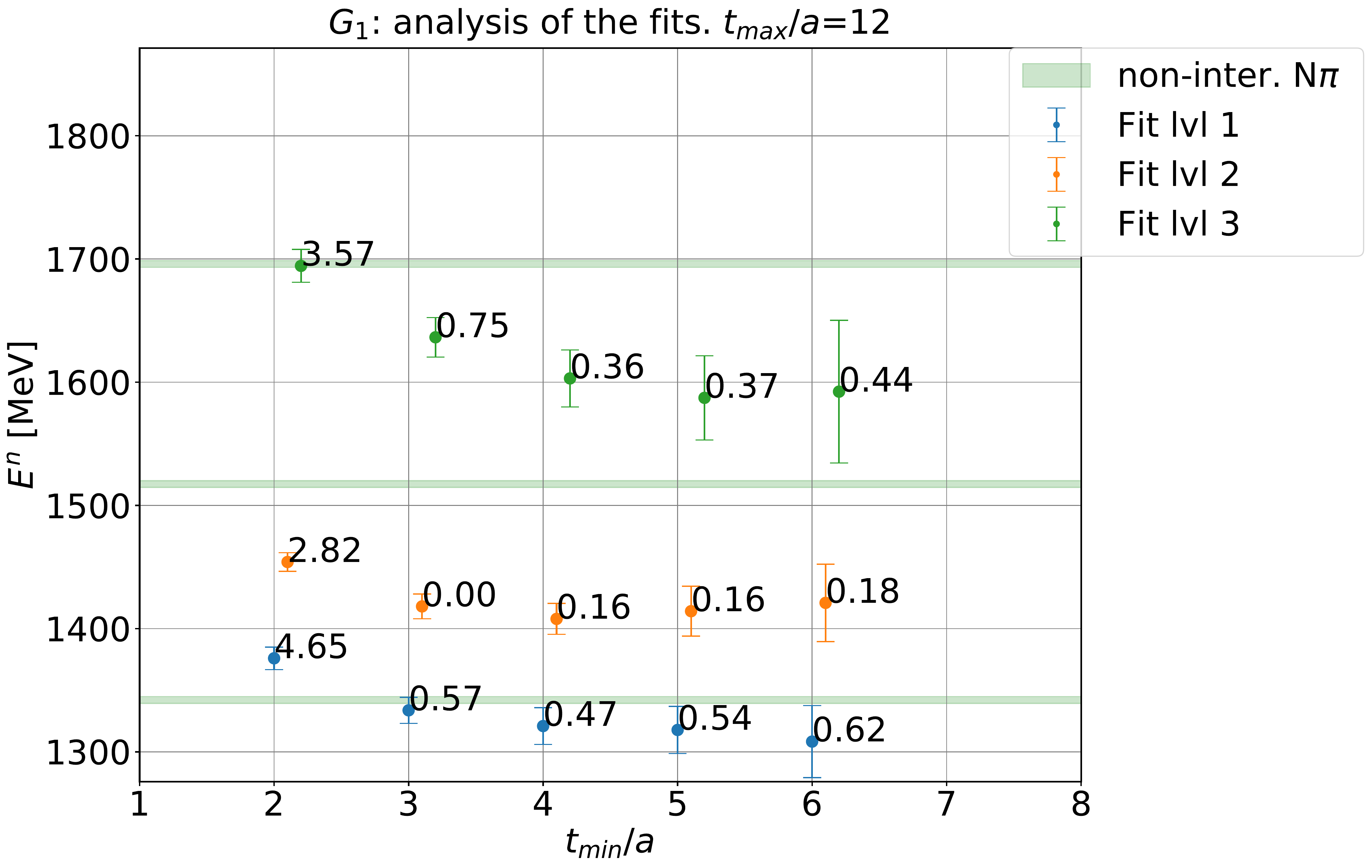}
\caption{\label{Fig:stfit} Stability analysis of the fits. The values of $\frac{\chi^2}{dof}$ are shown for each fit. Left panel: Rest frame $\vec{P}=\frac{2\pi}{L}(0,0,0)$: group $O_h^D$, irreps $H_g$ (forward-propagating) and $H_u$ (backward-propagating). The ground state shows a plateau starting at $t_{min}/a=5$ and the first excited at $t_{min}/a=4$. 
Right panel: Moving frame $\vec{P}=\frac{2\pi}{L}(0,0,1)$: group $C_{4v}^D$, irrep $G_1$. For all levels, $t_{min}/a=4$ has been chosen. }
\end{figure}

To ensure that early-time excited-state contamination is negligible, we perform stability tests by looking for a plateau while varying  $t_{min}$ in the fit time interval $\left[ t_{min},\,t_{max} \right]$. This is illustrated in Fig.~\ref{Fig:stfit}.

%% file: ScatteringAnalysis.tex
\section{Scattering analysis}
The L{\"u}scher quantization condition \cite{Briceno:2017max}, which connects the finite-volume energy levels with infinite-volume scattering phase shifts of two particles, is given by

\begin{equation}
{\rm det} \bigg( \mathbbm{1} + i t_{\ell}(s)(\mathbbm{1} + i {M}^{\vec{P}}) \bigg) = 0,
\end{equation}
where  $t_{\ell}(s) = \frac{1}{\cot{\delta_{\ell}(s)}-i}.$ Using the $t$-matrix parametrization and the total-angular-momentum basis of the interacting particles, we obtain
\begin{equation}
\label{eq:QC}
\det[M^{\vec{P}}_{Jlm,J^{\prime}l^{\prime}m^{\prime}}-\mathbold{\delta_{JJ^{\prime}}}\mathbold{\delta_{ll^{\prime}}}\mathbold{\delta_{mm^{\prime}}}\cot\delta_{Jl}]=0,
\end{equation}
where $M^{\vec{P}}_{Jlm,J^{\prime}l^{\prime}m^{\prime}}$ contains the finite-volume-spectra for the scattering of two particles with spins $\vec{\rm\bf s_1}$ and $\vec{\rm\bf s_2}$ and total linear momentum $\vec{P}$. The angular momentum $\vec{\rm \bf l}$ is the contribution from the $l$th partial wave such that the total angular momentum $\vec{\rm \bf J}$ is equal to $\vec{\rm \bf l} + \vec{\rm \bf s_1} + \vec{\rm \bf s_2}$. For fixed $J$ and $l$, we have $-J\leq m\leq J$. 
Thus, the $M$ matrix represents the mixing of the different angular momenta in finite volume. In the $t$-matrix,
$\delta_{Jl}$ denotes the infinit-volume phase shift for a given $J$ and $l$. It becomes quite evident from the addition of angular momenta that
$M$ is an infinite-dimensional matrix, because of the infinitely many possible values of $l$. 
To enable a lattice calculation, we need to select a cut-off $l_{max}$, 
and ignore higher partial waves. This can be justified for small center-of-mass momenta $p^*$, because $\delta(p^*)\propto (p^*)^{2l+1}$.

After neglecting higher partial waves, the finite-dimensional matrix $M$ can be further simplified into a block diagonal form through a basis transformation of the irreps of the symmetry groups of the lattice. Given a lattice symmetry group $G$ with irrep $\Lambda$ (from Table.~\ref{plan}), the matrix element in the new basis can be written as
\begin{equation}
\langle \Lambda r Jln \,|\, \hat{M}^{\vec{P}} \,|\, \Lambda^\prime r^\prime
J^\prime l^\prime n^\prime \rangle 
= 
\sum_{m,\, m^\prime}c_{Jlm}^{\,\Lambda r n} \,\,
c^{\,\Lambda^\prime r^\prime n^\prime}_{J^\prime l^\prime
  m^\prime}\,\,M^{\vec{P}}_{Jlm,J^\prime l^\prime m^\prime},
\end{equation}
where the row $r$ runs from $1$ to the dimension of $\Lambda$, $n$ labels the occurrence of the irrep, and $c_{Jlm}^{\,G \Lambda n}$ and $c^{\,G^\prime \Lambda^\prime n^\prime}_{J^\prime l^\prime m^\prime}$ are the relevant Clebsh-Gordan coefficients as calculated in \cite{G_ckeler_2012}.

\begin{figure}
\centering
\includegraphics[width=.7\textwidth]{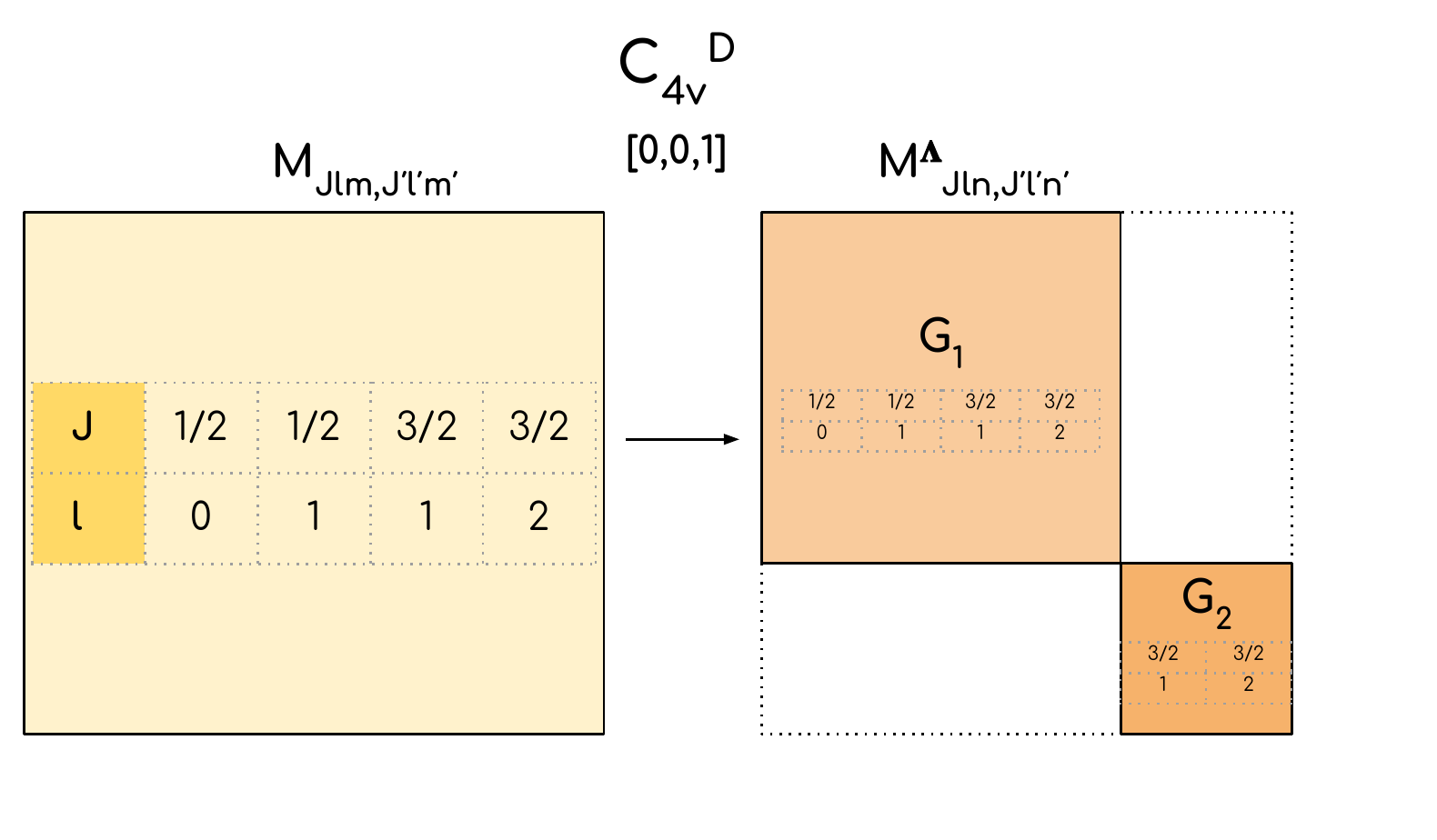}
\caption{ Block diagonalization of the $M$ matrix in the irrep basis. The case shown is for the $C^D_{4v}$ group. }
\label{Fig:qcc4v}
\end{figure}

From Schur's lemma, we know that $\hat{M}^{\vec{P}}$ is block-diagonalized in the new basis,
\begin{equation}
\langle \Lambda r Jln \,|\, \hat{M}^{\vec{P}} \,|\, \Lambda^\prime r^\prime
J^\prime l^\prime n^\prime \rangle 
= \delta_{\Lambda \Lambda^\prime} \delta_{r r^\prime} M^{\Lambda}_{Jln,J^\prime l^\prime n^\prime}.
\end{equation}
The advantage of the basis transformation can be observed for example in the moving frame $\vec{P} = \frac{2\pi}{L}(0,0,1)$ 
with symmetry group $C^D_{4v}$. With two irreps $G_1$ and $G_2$, Eq.~(\ref{eq:QC}) simplifies to 
one quantization condition per irrep
\begin{equation}
\label{eq:g1}
{\rm det}( M^{G_1}_{Jln, J'l'n'} - \delta_{JJ'}\delta_{ll'}\delta_{nn'}\cot{\delta^{G_1}_{Jl}}) = 0,
\end{equation}
\begin{equation}
\label{eq:g2}
{\rm det}( M^{G_2}_{Jln, J'l'n'} - \delta_{JJ'}\delta_{ll'}\delta_{nn'}\cot{\delta^{G_2}_{Jl}}) = 0.
\end{equation}
This basis transformation, along with the angular-momentum content, is also shown schematically in Fig.~\ref{Fig:qcc4v}. 
The angular-momentum content depicts the mixing of various partial waves which contribute to the $J=3/2$ channel.

The quantization conditions  can be written in terms of $w_{js}$ functions( where $|l-l'|\leq j\leq l+l'$ and $-j\leq s\leq j$) that include the generalized Zeta functions as in Ref.~\cite{G_ckeler_2012}. For the case of $G_2$, considering all relevant $J,l$ values of the phase shift $\delta_{J,l}$ present in the energy region of interest, the quantization condition can be written as
\begin{equation}
9 \left(w_{1,0}-w_{3,0}\right){}^2-25 \left(-\cot\delta_{\frac{3}{2},1}^{G_2}+w_{0,0}-w_{2,0}\right) \left(-\cot\delta_{\frac{3}{2},2}^{G_2}+w_{0,0}-w_{2,0}\right)=0.
\label{eq:mixing}
\end{equation}

The mixing between $l=1$ and $l=2$, is clearly shown in Eq.~(\ref{eq:mixing}). If we neglect contributions from $l=2$,
we arrive at the quantization condition as used in Ref.~\cite{G_ckeler_2012}.

Due to the existence of eigenstates with  same total angular momentum $J$, but different orbital angular momenta $l$, we have mixtures in scattering amplitudes originating from nearby $\Delta$ resonances (see Tab~\ref{resonances}). In this preliminary study, we are only interested in the phase shift $\delta_{J=\frac{3}{2},l=1}$. Having fixed $l_{max}$, we neglect the contributions from higher partial waves, but we need to take into account
contributions from lower partial waves and $J$'s for the complete analysis.

%% file: Phase_shift_results.tex
\section{Phase-shift results}

\begin{figure}[!htb]
\centering
\includegraphics[width=7.5cm]{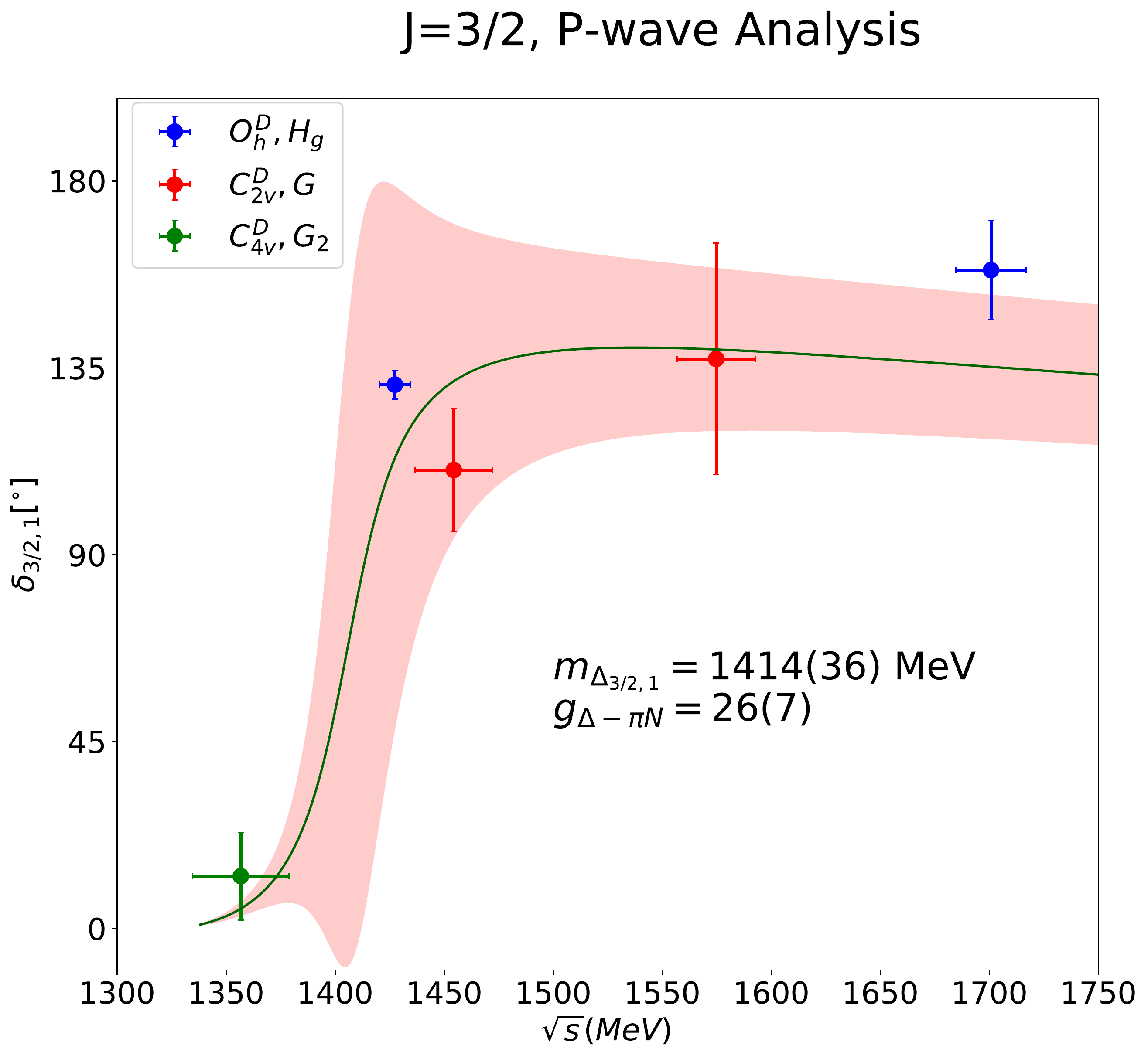}
\caption{Preliminary results for the $J=3/2$, $P$-wave, $N$-$\pi$ scattering phase shift.}
\label{phase}
\end{figure}

Our preliminary results for the scattering phase shifts obtained from the L\"uscher quantization condition are shown in Fig.~\ref{phase}.
To extract the resonance parameters, we directly fit a model for the $t$-matrix to the energy spectra; this methodology is known as the inverse L\"uscher analysis.

The scattering amplitude for a narrow resonance in QCD can be characterized by the Breit Wigner parametrization,
\begin{equation}
\label{eq:breit}
t_{\ell}(s) = \frac{\sqrt{s}\,\Gamma(s)}{m_R^2 - s - i \sqrt{s}\,\Gamma(s)},
\end{equation}
where $s$ is the center of mass energy squared, $m_R$ is the mass of resonance, and $\Gamma{(s)}$ is the decay width of the resonance.
For the $\Delta$ resonance, the decay width can be expressed in effective field theory \cite{Pascalutsa:2005vq} to lowest order as
\begin{equation}
\label{eq:decay}
\Gamma^{LO}_{EFT}= \frac{ g_{\Delta-\pi N}^2}{48 \pi}\frac{E_N + m_N}{E_N + E_\pi}\frac{p^{*3}}{m_N^2},
\end{equation}
with the dimensionless coupling $g_{\Delta-\pi N}$ and center-of-mass momentum $p^*$.

We denote the lattice result for the $n^{th}$ finite-volume energy level in the $\Lambda$ irrep of the moving frame with momentum $\vec{P}$ as $\sqrt{s_n^{\Lambda, \vec{P}}}^{[avg]}$. The corresponding model function $\sqrt{s_n^{\Lambda, \vec{P}}}^{[model]}$ is constructed by combining the L\"uscher quantization condition
(\ref{eq:QC}) with the Breit-Wigner model given by Eqs.~(\ref{eq:breit}) and (\ref{eq:decay}). We define the $\chi^2$ function to be minimized
as
 \begin{equation}
 \label{eq:chi2}
\chi^2 = \sum_{\vec{P},\Lambda,n} \sum_{\vec{P}^\prime,\Lambda^\prime,n^\prime} \left( \sqrt{s_n^{\Lambda, \vec{P}}}^{[avg]}\
	 - \sqrt{s_n^{\Lambda, \vec{P}}}^{[model]}\right) [C^{-1}]_{\vec{P},\Lambda,n;\vec{P}',\Lambda',n'}
	\left( \sqrt{s_{n'}^{\Lambda', \vec{P}'}}^{[avg]} 
	- \sqrt{s_{n'}^{\Lambda', \vec{P}'}}^{[model]}  \right).
\end{equation}
We include all irreps of the two groups $O_h^D$ and $C_{2v}^D$ and one irrep of $C_{4v}^D$. The fit gives
\begin{eqnarray}
 m_\Delta &=& 1414 \pm 36\:\:{\rm MeV}, \\
  g_{\Delta-\pi N} &=&26 \pm 7,\\
  \dfrac{\chi^2}{dof} &=&1.05,
\end{eqnarray}
and the corresponding phase shift curve is also shown in Fig.~\ref{phase}.

%% file: Discussion.tex
\section{Discussion}

We compare our preliminary results for $m_\Delta$ and $g_{\Delta-\pi N}$ with other lattice calculations and
with the values extracted from experiment in Table \ref{contemp}. Our results are consistent with previous
lattice determinations using the L\"uscher method at similar pion masses.

This preliminary analysis was done with only $1/3$ of the total number of configurations available for the \texttt{A7} ensemble, and
we are now increasing the statistics. Furthermore, we plan to add the \texttt{A8} ensemble (Table \ref{tab:lattice}) in the near future.
The phase-shift curve will be constrained more precisely with the additional points calculated from the bigger $(32^3 \times 48)$ lattice at
the same pion mass. We also plan to include non-resonant contributions from other possible $J$'s and $l$'s, taking into account partial-wave mixing.
Finally, we will investigate other decay-width models to better understand the model-dependence of the extracted resonance parameters.
 
\begin{table}[h]
\resizebox{\textwidth}{!}{%
\begin{tabular}{@{}lllll@{}}
\toprule
Collaboration                                      & $m_\pi$ [MeV]    & Methodology                    & $m_\Delta$ [MeV] & $g_{\Delta-\pi N}$ \\
\midrule
Verduci (2014) \cite{Verduci:2014btc}              & 266(3)        & Distillation, L\"uscher        & 1396(19)      & 19.90(83) \\
Alexandrou et.al. (2013) \cite{Alexandrou:2013ata} & 360             & Michael, McNeile               & 1535(25)      & 26.7(0.6)(1.4) \\
Alexandrou et.al. (2015) \cite{Alexandrou:2015hxa} & 180             & Michael, McNeile               & 1350(50)      & 23.7(0.7)(1.1) \\
Andersen et.al. (2017) \cite{Andersen:2017una}     & 280             & Stoch.~distillation, L\"uscher & 1344(20)      & 37.1(9.2) \\
Our result (preliminary)                           & 258.3(1.1)    & Smeared sources, L\"uscher           & 1414(36)     & 26(7) \\
\hline
Physical value \cite{Olive:2016xmw}                & 139.57018(35) & phenomenology, K-matrix               & 1232(1)       & 29.4(3), 28.6(3) \\
\hline
\end{tabular}%
}

\caption{Comparison of results for $m_\Delta$ and $g_{\Delta-\pi N}$.}
\label{contemp}
\end{table}

%% file: Acknowledgements.tex
\section{Acknowledgements}
This research used resources of the National Energy Research Scientific Computing Center (NERSC), a U.S. Department of Energy Office of Science User Facility operated under Contract No.~DE-AC02-05CH11231. SM and GR are supported by the U.S. Department of Energy, Office of Science, Office of High Energy Physics under Award Number DE-SC0009913. SK is supported by the Deutsche Forschungsgemeinschaft grant SFB-TRR 55. SK and GS were partially funded by the IVF of the HGF. SM and SS also thank the RIKEN BNL Research Center for support. JN was supported in part by the DOE Office of Nuclear Physics under grant DE{}-SC-0{}011090. AP was supported in part by the U.S. Department of Energy Office of Nuclear Physics under grant DE-{}FC02-06ER41444. LL was supported by the U.S. Department of Energy, Office of Science, Office of Nuclear Physics under contract DE-AC05-06OR23177. SP is supported by the Horizon 2020 of the European Commission research and innovation programme under the Marie Sklodowska-Curie grant agreement No.~642069. We acknowledge the use of the USQCD software QLUA for the calculation of the correlators.